\documentclass[preprint,journal]{vgtc}       

\ifpdf
  \pdfoutput=1\relax                   
  \usepackage{graphicx}                
  \DeclareGraphicsExtensions{.pdf,.png,.jpg,.jpeg} 
\else
  \usepackage{graphicx}                
  \DeclareGraphicsExtensions{.eps}     
\fi%

\graphicspath{{figures/}{pictures/}{images/}{./}} 
\usepackage{paralist}
\usepackage{microtype}                 
\PassOptionsToPackage{warn}{textcomp}  
\usepackage{textcomp}                  
\usepackage{mathptmx}                  
\usepackage{times}                     
\usepackage{cite}                      
\usepackage{tabu}                      
\usepackage{booktabs}                  
\usepackage{balance}

\onlineid{1329}

\vgtccategory{Research}
\vgtcpapertype{application/design study}

\title{RISeer: Inspecting the Status and Dynamics of Regional Industrial Structure via Visual Analytics}

\author{Longfei Chen, Yang Ouyang, Haipeng Zhang, Suting Hong, and Quan Li}
\authorfooter{
\item
 L. Chen and Y. Ouyang are with School of Information Science and Technology, ShanghaiTech University. E-mail: chenlf,ouyy@shanghaitech.edu.cn. They contributed equally to this work.
\item
 S. Hong is with School of Entrepreneurship and Management, ShanghaiTech University. E-mail: hongst@shanghaitech.edu.cn.
\item
 H. Zhang and Q. Li are with School of Information Science and Technology, ShanghaiTech University. Quan Li is the corresponding author. E-mail: zhanghp,liquan@@shanghaitech.edu.cn.
}

\shortauthortitle{Biv \MakeLowercase{\textit{et al.}}: Global Illumination for Fun and Profit}

\abstract{
Restructuring the regional industrial structure (RIS) has the potential to halt economic recession and achieve revitalization. Understanding the current status and dynamics of RIS will greatly assist in studying and evaluating the current industrial structure. Previous studies have focused on qualitative and quantitative research to rationalize RIS from a macroscopic perspective. Although recent studies have traced information at the industrial enterprise level to complement existing research from a micro perspective, the ambiguity of the underlying variables contributing to the industrial sector and its composition, the dynamic nature, and the large number of multivariant features of RIS records have obscured a deep and fine-grained understanding of RIS. To this end, we propose an interactive visualization system, \textit{RISeer}, which is based on interpretable machine learning models and enhanced visualizations designed to identify the evolutionary patterns of the RIS and facilitate inter-regional inspection and comparison. Two case studies confirm the effectiveness of our approach, and feedback from experts indicates that \textit{RISeer} helps them to gain a fine-grained understanding of the dynamics and evolution of the RIS.
} 

\keywords{Spatiotemporal dynamics, multivariate time series, regional industrial structure, visualization}

\CCScatlist{ 
 \CCScat{K.6.1}{Management of Computing and Information Systems}%
{Project and People Management}{Life Cycle};
 \CCScat{K.7.m}{The Computing Profession}{Miscellaneous}{Ethics}
}

\vgtcinsertpkg

\begin{document}


\firstsection{Introduction}

\maketitle

\par Regional industrial structure (RIS) refers to the proportional relationship between industrial sectors with different development functions in a specific region, and it is the result of the combination of the spatial layout of the national economy in a specific region~\cite{aleksander1979great}. In other words, the reason why a certain industrial structure emerges is determined both by the advantages of a particular region and by the overall requirements of the spatial layout of the national economy. Adjusting the RIS is fundamental to curbing the economic downturn and achieving economic revitalization. Various empirical evidences show that the economic development of regions around the world is always linked to the optimization of industrial structure~\cite{kai2003study,peneder2003industrial,zhu2019exploring}. For example, the relative decline of the United States and the rise of Japan in the late $1970$s and early $1980$s, as well as the late recovery of the United States and the decline of Japan in the $1990$s, can be attributed to a concerted restructuring of regional industries.

\par Understanding the current status and dynamics of the RIS will greatly help to examine and assess the corresponding rationality of the industrial structure and the benefits to stakeholders. It is worth noting that for entrepreneurs, identifying an appropriate timing to enter the market during the dynamic change of the RIS is crucial. For investors, an objective evaluation of the current RIS can provide insight into the allocation of rational upstream or downstream industries. For regional planners and policy makers, an examination of the dynamics of the RIS can help them make feasible assessments, develop regional planning strategies, and attract more potential investment opportunities. Given the characteristics of RIS, there is therefore an urgent need for a methodology that can adequately reflect and examine the current situation and dynamics of RIS. Previous studies on RIS have mainly focused on qualitative and quantitative research to analyze and adjust the rationality from a macro perspective. Generally speaking, most studies start from the changes of industrial structure and national policies, and conduct statistical analysis of historical data. Then, with the help of econometric models, they estimate the evolutionary trends of RIS. Some studies use information on regional industrial enterprises and production resources to study the dynamic proportional relationships between industries~\cite{du2019assessing,evseev2020digital,lopatin2019methodological,materna2019acquisition}. Clustering and comparative analysis was also performed based on the industrial sectors corresponding to enterprises registered in a specific region~\cite{abramov2016regional,feser2008clusters,liu2018big}, which performed well in understanding RIS dynamics.

\par This work focuses on tracing the current status and dynamics of the RIS consisting of and reflecting the enterprise activities behind it~\cite{atack1986firm,chen2006development,drucker2012regional,knoben2016agglomeration}. This can complement the analysis of RIS from a microscopic perspective. However, it still poses several challenges. First, industrialization and modernization have enabled countless businesses to be established, survive and grow. During this process, profile information on newly established businesses and all corresponding major industrial and commercial changes are recorded. Traditional methods usually aggregate the proportions of the various industries in a given region and observe trends over time. However, they cannot assess whether and how the underlying variables affect these trends. While some works consider correlations between different industrial sectors, they still do not directly show the relationships between input characteristics and output ratios of industries, as well as potential problems, such as anomalies. Second, while most previous works are beneficial for monitoring the evolutionary patterns of RIS, fine-grained analyses such as industrial agglomeration, radiation and migration, which are often hidden behind large-scale business activities, are challenging. An effective approach should provide clear visual cues to capture potential RIS patterns, identify anomalous changes, and make cross-sectional and regional comparisons. Third, archival records of enterprises are inherently multi-attribute items with a variety of data types, such as timestamp, text, numeric, and geographical pairs, which together determine where a business belongs in a given industrial structure. More and more enterprises are being registered and their information is being added/updated over time. Given the long lead time of industrial structure studies, capturing the impact of these attributes on the dispersion and temporal changes of RIS requires careful visualization.

\par In this study, we propose a visual analytics approach to understanding the current status and dynamics of RIS from the perspective of industrial enterprise-level information. In particular, to address the first challenge, we treat the dynamics of the enterprise establishments as a regression problem and use multiple machine learning predictive models to explain potential associations between input features and output RIS patterns as well as possible anomalies. To address the second and third challenges, we employ an enhanced timeline design with novel glyphs to encode the dynamics of RIS in geographical space over time and reveal the impact of temporal changes in multiple attributes on the region at different time periods. We propose \textit{RISeer}, an interactive visualization system that identifies the evolutionary patterns of RIS and facilitates inter-regional comparisons based on visual designs. We demonstrate the usefulness of \textit{RISeer} using real-world datasets and case studies. In addition, we collected feedback on our approach from a number of domain experts. Our main contributions are: 1) We use information at the industrial enterprise-level to characterize these issues in the context of studying the current state and dynamics of the RIS through thorough discussion of design requirements with domain experts; 2) We develop interactive visualizations with new visual features to support the study of the current state and dynamics of RIS. To the best of our knowledge, \textit{RISeer} is the first endeavor of this kind.

\section{Related Work}
\subsection{Regional Industrial Structure Analysis}
\par Previous studies on regional structural analysis have focused on the rationalization and adjustment of the RIS. Notably, they would use econometric models to estimate the evolutionary trend of RIS~\cite{li2019demand,peneder2003industrial,yu2018impact} or conduct qualitative and quantitative analysis from a macro perspective~\cite{kai2003study,townroe1991rationality,zhu2019exploring}. For example, Lu et al.~\cite{lu2008empirical} used a panel data model to measure the contribution of different industries to economic growth and concluded that economic growth is highly correlated with the development of secondary and tertiary industries. Zhu et al.~\cite{huaxiong2011research} used a dynamic shift-share model to study the different regional industry components with different characteristics. These works combined with historical data and applied economic methods to analyze and summarize some patterns of RIS. However, the analysis of some key dimensions, such as the spatio-temporal dimension, is still missing. Visualization techniques have also been introduced to help analyze industrial structure. For instance, Zhang et al.~\cite{zhang2018citespace} applied \textit{CiteSpace} to study industrial restructuring-related literature in the Chinese Social Sciences Citation Index (CSSCI) database between $1998$ and $2017$ by mapping the visual knowledge of high-impact authors, research institutions, and co-occurring keywords. Jiang~\cite{zilong2014analysis} revealed regional development patterns by visualizing agglomeration and equilibrium evolution characteristics and regional differences. Zhou et al.~\cite{zhou2018visual} visualized spatiotemporal and multivariate GDP data to capture the dynamic characteristics of economic developments. Later, they further utilized input-output table (IOT) data to measure industry correlations and help users identify relevant temporal changes in the underlying IOT data. Their goal is to combine the extraction of economic clusters over a time period with the tracking of the evolution of dynamic features across time. The characteristics of different clusters are projected through \textit{MDS}, and the proposed visualization methods similar to Sankey diagram~\cite{riehmann2005interactive} allow analysts to visualize the dynamics of the clusters. However, their work examines GDP data on a provincial basis, and their visualization does not provide a good depiction of the internal structure of spatial clusters, making it difficult for analysts to access the data in a comprehensive way. In this work, we study the current state and dynamics of RIS from the perspective of fine-grained business registration data through elaborate visualization, which inherently influences and depicts the evolution of industrial structure in a given region.

\subsection{Time-Series Prediction and Evaluation}
\par Forecasting the number of registered enterprises over a period of time belongs to the category of time series forecasting. In the past decades, various time series forecasting methods have been proposed, including classical statistical methods and machine learning methods, aiming to predict future trends based on empirical and data~\cite{10.1145/3411764.3445083}. For example, different time series forecasting models based on univariate time-series have been proposed, such as auto-regression-based methods~\cite{box2015time,stellwagen2013arima} and exponential smoothing (ES) methods~\cite{hyndman2008forecasting}. Later, these traditional methods were extended to handle multivariate time series~\cite{johansen1995likelihood} considering relevant properties. Vector Error Correction Model (VECM) was proposed to predict the stock market~\cite{mukherjee1995dynamic}, and Random Forest (RF) was applied to predict H5N1 avian influenza outbreaks~\cite{kane2014comparison}. With the development of deep learning and its variants, advanced models such as recurrent neural networks (RNNs)~\cite{hochreiter1997long} have been applied to solve multivariate time series forecasting. In this work, we have implemented some of the most representative algorithms of different classes in the scenario of industrial structure analysis. Also, various methods have been proposed to evaluate and compare different time series forecasting models~\cite{armstrong1992error,camastra2009comparative,hyndman2006another,khair2017forecasting}. We used mean absolute percentage error (MAPE) as a measure to evaluate our multivariate forecasting models because it overcomes the scale dependence~\cite{barrow2010evaluation,sharda1992connectionist}. We further provide an explanation for the comparison and evaluation of RIS by revealing the importance of different features in the forecasting process. 

\subsection{Visualization of Spatiotemporal Data}
\par Visual analysis of information related to regional enterprises, such as location, belongs to the visualization of spatiotemporal data with multiple attributes. In particular, when analyzing the evolutionary dynamics of RIS, well-designed visualizations are needed to integrate the spatiotemporal aspects of the data. For example, Andrienko et al.~\cite{andrienko2003exploratory} surveyed the literature on existing experimental techniques related to spatiotemporal data and corresponding analysis tasks. In general, spatial data distributions are typically represented using heat maps~\cite{hilton2011saferoadmaps,maciejewski2009visual,mehler2006spatial}, choropleth maps~\cite{stein2016game}, and glyph-based visualization techniques~\cite{10.2312:conf:EG2013:stars:039-063} to exploit, which are intuitive for users to perceive and understand spatial data features. However, location data, usually represented by pins on the map, will inevitably hide temporal information. Efforts such as color-coding different times of the day~\cite{fuchs2013evaluation} and drawing connecting lines or curves through spatiotemporal cubes indicate connectivity. However, they can introduce visual clutter and occlusion, making the results quite challenging to perceive. The use of animations~\cite{li2018embeddingvis} also limits overview capabilities and introduces overloaded thinking. In this work, we divide the temporal evolution of the enterprises into multiple parts and visualize the aggregated enterprises as elaborate glyphs based on their aggregated geographical locations. We then use connected paths to model the evolutionary dynamics of the RIS.

\section{Observational Study}
\subsection{Background}
\par In order to understand how (well) domain experts analyze in practice the current state and dynamics of RIS, we worked with a team of experts from a partner university and two governmental functions. They were two researchers specializing in innovation-oriented entrepreneurship and fintech (E1, female, age: $33$; E2, male, age: $34$), a business administrator (E3, male, age: $38$), and a local industrial park planner (E4, female, age: $31$). Their research and work have significant overlap in providing insight into local economic markets through qualitative and quantitative analysis of information on business and industry activities~\cite{delanoe2013intention,hong2018dynamic,hong2020competition}. For example, E1 uses theoretical models and empirical analysis to examine the different forces that influence the development of enterprises and startups, and ultimately the RIS. She also mentioned an effort proposed by the Harvard Growth Lab\footnote{https://growthlab.cid.harvard.edu/videos/introducing-metroverse-growth-labs-urban-economy-navigator} called \textit{Metroverse}, which aims to provide stakeholders with unprecedented economic data on more than $1,000$ cities around the world to help identify pathways to growth and diversification. She likes the idea of using visualizations to bring research to the city level and answer questions such as ``\textit{what is the economic makeup of a city and where does that city fit in the industrial space?}'' E2, on the other hand, focuses on using advanced automated methods to predict future trends in business and the accompanying industrial structure in terms of proportions in each industry and cross-industry transformation. Based on their findings and recommendations, E3 and E4 will work together to assess the current state of RIS in conjunction with a spreadsheet of relevant businesses and startups registered with local government functions to infer key elements of RIS dynamics to gain insights and develop strategies accordingly.

\par Nevertheless, in analyzing RIS dynamics, the experts had some concerns. First, E3 and E4 commented, ``\textit{we are not sensitive to updates in information about registered companies.}'' Given the nature of the data, experts would like to have some visual cues about potential patterns of interest in the historical data and possible future trends for further decision-making needs. Second, in exploring information related to enterprise registration, E1 and E2 attempted to plot all enterprises on a geographical map based on latitude and longitude coordinates, which created significant visual confusion. ``\textit{In addition to the scale, the points may have shapes that represent categories of enterprises and colors that show another attribute, making them difficult to present, let alone understand,}'' said E.2. In addition, these records exhibit spatiotemporal variation, which ``\textit{is not easy to address with a general summary based on statistics.}'' Experts wondered if we could visualize these records in a more abstract way, while maintaining details of dynamic and multivariate attributes. Third, current practices focuses on RIS from a global and high level perspective. For example, E3 and E4 often examine RIS dynamics from the perspective of sectoral proportions and cross-sectoral shifts, which is shallow. E4 shared one of his observations about the analysis of agglomeration-performance relationship in certain regions: the initial goal of an industrial park or region is to target a specific industry and integrate upstream and downstream enterprises in the chain to form spatial aggregation of property rights for all enterprises in the industrial chain. However, most regional industrial parks encounter a similar problem: ``\textit{Most industrial parks, due to initial financial pressures, eventually drift into a spatial aggregation of property in different and unrelated industries}''. Thus, experts were curious to perform a retrospective analysis of past and existing RIS in an attempt to answer questions such as ``\textit{does regional industrial agglomeration really benefit enterprises?}'' Thus, a nontrivial visualization is needed to clarify the ``agglomeration-performance'' relationship behind enterprise activities.

\begin{figure*}[h]
    \centering
    \includegraphics[width=\textwidth]{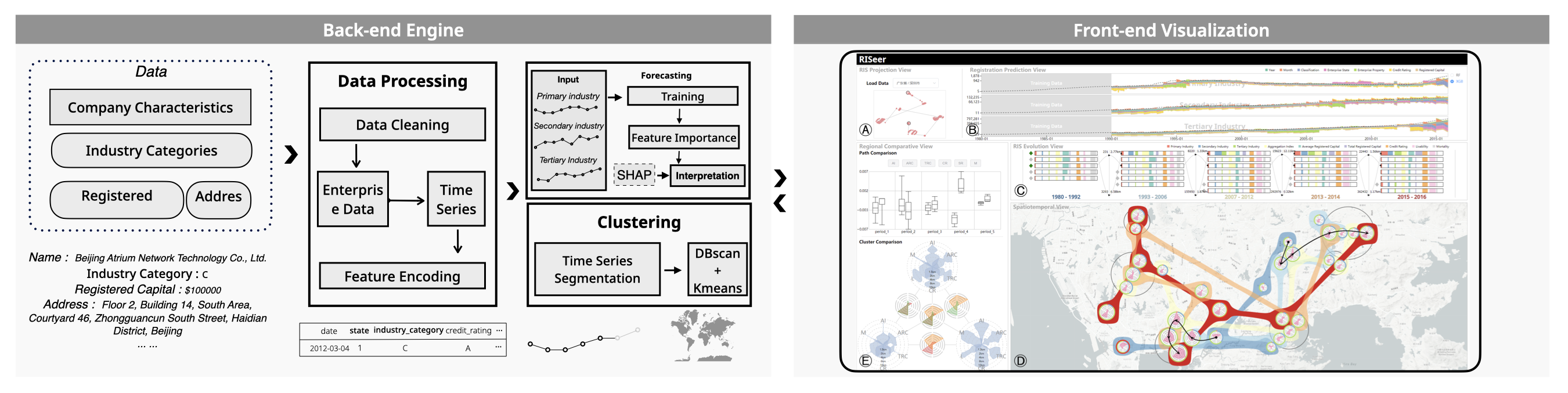}
    \vspace{-6mm}
    \caption{\textit{RISeer} pipeline. (A) RIS projection view obtains the overall distribution of RIS patterns over the entire periods; (B) Registration prediction view predicts the number of business registration and reveal feature importance; (C) RIS evolution view explores regions' characteristics and evolution over time; (D) Spatiotemporal view projects the identified regional clusters onto the map. (E) Box plots of growth rate of business registration activity.}
    \label{fig:pipeline}
    \vspace{-6mm}
\end{figure*}

\subsection{Experts' Needs and Expectations}
\par We interviewed E1 -- E4 to identify their concerns and potential barriers to using business and industry information to effectively analyze RIS dynamics. At the end of the interviews, the need for a visualization system was a key theme in the feedback gathered. We followed a user-centered design framework using discussion, brainstorming, and design. After several iterations, we gathered their input and condensed it into a set of following requirements. The three \textbf{Industrial Sector-Level} requirements \textbf{(R.1 -- R.3)} provide stakeholders with a general sense of the dynamics of the industrial sectors and the three \textbf{Geospatial Region-Level} requirements \textbf{(R.4 -- R.6)} enable stakeholders to understand the aggregation, migration, and evolution of industrial structures over time within a given geospatial region.

\par \textbf{R.1 Explain the temporal dynamics in the industrial and business sectors.} The traditional approach is to transfer the change in information from the enterprise level to its corresponding sector level by calculating its contribution to the sector. While understanding the change in ratios for each sector provides an overview of the dynamics of all sectors over time, it only considers absolute numbers and ignores factors that may influence change in RIS dynamics. E1 is interested in explaining specific increases, outbreaks, or decreases on the time axis. Therefore, equipping the system with factor analysis allows to observe and speculate on the factors that influence the RIS temporal dynamics.

\par \textbf{R.2 Explore general patterns and potential outliers in the dynamics of sector change.} Experts need to quickly browse through large amounts of enterprise data to identify areas of interest. For example, sectors with unprecedented growth are more likely to attract the expert's attention. Further investigation is needed to explain general patterns and potential outliers across historical periods. For example, E1 -- 2 would like to find out if there are standard characteristics and outliers between industrial sectors and the possible reasons behind them.

\par \textbf{R.3 Predict the future composition of the RIS and explore the influencing factors behind it.} Another conventional approach to RIS analysis is to understand the composition of the industrial sector, for example, the ratio and the appropriate number of establishments. Based on the history and current status of RIS, experts want to know the future trends and influencing factors of the quality of the composition of each industrial sector. Thus, they can make speculations and assumptions about the future before developing the corresponding strategies.

\par \textbf{R.4 Summarize the dynamics and multivariance of enterprises.} As mentioned earlier, directly mapping millions of enterprises on a geographic map without any abstraction or simplification is bound to create visual clutter, compromise effective understanding, and affect the performance of the system. In addition, as time goes on, more and more new businesses are being established and added to the existing environment. Integrating these factors requires careful design to summarize the dynamics and multiple attributes of large-scale enterprises.

\par \textbf{R.5 Reveal the regional ``agglomeration-performance'' relationship resided in various groups of enterprises.} According to E3 -- 4, enterprise-level performance may be strongly related to the geospatial locations of agglomerations. Although theoretical studies have shown predominantly on positive performance effects as an incentive for enterprise collocation ``\textit{to explain the emergence of agglomerations}''~\cite{bell2005clusters,molina2003impact}, experts have also focused on adverse performance effects, ``\textit{some enterprises may benefit from agglomeration, while others may be harmed by aggregation and relocate elsewhere,}'' said E4. Thus, in the face of these conflicting empirical findings, the net performance effects of enterprises located in geographic agglomerations remain ambiguous~\cite{knoben2016agglomeration,myles2000agglomeration,sorensen2003conception}, calling for clarification of, e.g., ``the agglomeration-performance'' relationship hidden in the large-scale activities of enterprises.

\par \textbf{R.6 Compare and track performance at the region level.} Having summarized the dynamics and multiple attributes of enterprise at the regional level, experts want to conduct in-depth analyses of RIS, such as determining regional performance, comparing multiple regions at a specific point in time, and tracking the evolution of a particular region over a long period of time. For example, comparing two regions at a specific timestamp makes it possible to identify differences in their positioning, while tracking their performance over time can facilitate research into the reasons behind booms or busts.

\section{Approach Overview}
\par We propose \textit{RISeer}, an interactive visual analytics system that helps to understand the current status and dynamics of the RIS from the perspective of industrial enterprise registration information. \autoref{fig:pipeline} shows the architecture of \textit{RISeer}, which consists of a back-end engine in terms of a data processing module and a forecasting module, and a front-end visualization. In the back-end engine, the features hidden in the original raw data should first be quantified and transformed into attributes acceptable to the time-series multivariate inputs in the data processing module before building the prediction model. In the prediction module, various machine learning models are first trained based on the input data and features, and then these models are employed to make registration predictions. In addition, we determine the importance of the features for each model. We also divide the business registration data into segments and generate clusters for each part. In the front-end visualization, We provide several elaborate visual designs for analysis.

\section{Back-end Engine}
\subsection{Data Description and Processing}
\par The collaboration with the domain experts provided us with a publicly available dataset of business registration records collected from the National Enterprise Credit Information Publicity System\footnote{http://www.gsxt.gov.cn}, covering the period from $1980$ to $2015$. Each record contains information such as the enterprise's primary identification code, name, address, and registered capital. Based on suggestions from domain experts, we developed and studied the following attributes in this work: 1) address: it records detailed geographic information about a business; 2) start and end dates of operation: it allows us to track the the operations of businesses that existed during a specific period for further analysis; and 3) basic information, including industry category, credit rating, registered capital, and enterprise status, among others. This information can be used to describe a specific business.

\par To understand the evolutionary pattern of RIS, we transform the registration records of existing enterprises into a time-series dataset. Then, we use a multivariant time series prediction machine learning model to obtain the basic features of the evolutionary patterns. Forming time series data from raw business registration data is nontrivial due to the large volume and complex attributes of raw business registration data. Moreover, the temporal characteristics, i.e., the start and end dates of the business, are hidden in each record independently of each other. Therefore, considering the computational efficiency and speeding up the processing, we change the index of the data from the main identity code of the enterprise to the date and construct a hash table to reorganize the original information. Before modeling the time series data, several non-numerical dimensions, such as industry category and credit rating, must be quantified. Since these attributes have a limited and relatively fixed range of values, it is acceptable to map them to numbers in order to maintain the information encoded using one-hot encoding.

\subsection{Time-series Forecasting Machine Learning Models}
\par We define the input to a time series prediction machine learning model as a sequence of historical data $X_n = {x_{n-L}, x_{n-(L-1)}, ..., x_{n-1}, x_n}$, where $L$ is the parameter that is adapted for different datasets and $x_n$ is a multidimensional feature vector with timestamp $n$ (i.e., the feature vector consists of $7$ dimensions, including \textit{year}, \textit{month}, \textit{enterprise classification code}, \textit{registered capital}, \textit{credit rating}, \textit{enterprise property}, and \textit{enterprise state}) denoted as $x_n^f \in R$. The goal of a multivariate time series forecasting models is to predict specific values at some future time stamp. In our case, we provide only a single-step forecast with an output market of $y_n = x_{n+1}$. Considering that different countries divide the industrial structure in different ways, but basically divided into three main categories, we first divide the original dataset into three significant industries based on industry category, namely, \textit{primary industry}, \textit{secondary industry}, and \textit{tertiary industry}. Then, we calculate the number of enterprises existing in each month from $1980$ to $2015$. We use the data from $1980$ to $1990$ as the first training set to predict the trend in $1991$, followed by the data before $1991$ as the training set to predict the trend for $1992$, and so on. The reason for performing stepwise forecasting is based on the practical requirements of most real-world time series forecasting tasks. Thus, the final prediction curve is spelled out for each time unit. We use the classical \textit{MAPE} metric to evaluate the performance of the forecasting model. There are several representative time series forecasting models available for evaluation, such as \textit{Arima}~\cite{stellwagen2013arima}, \textit{Vector Auto-Regression Model (VAR)}~\cite{lutkepohl2013vector}, \textit{Random Forest Model (RF)} or \textit{Random Decision Forests}~\cite{ho1995random} and \textit{Long Short-Term Memory Recurrent Neural Networks (LSTM)}, covering linear, nonlinear regression methods, and machine learning models~\cite{10.1145/3411764.3445083}. In this work, we considered both model performance and model interpretability, and finally select two machine learning models for multivariate time series prediction, namely, RF and \textit{XGBoost}~\cite{chen2016xgboost}, as they have sufficiently high model accuracy and good interpretability.

\par To explore the dynamic evolution of the time-series-based prediction model for each business sector, we further provide model explanations to help users understand the model performance and its differences in terms of feature importance~\cite{10.1145/3411764.3445083}. Notably, we chose to use \textit{SHapley Additive exPlanations (SHAP)} values to check the correlation between the input feature importance and the output prediction results for business registrations~\cite{10.5555/3295222.3295230}. We chose \textit{SHAP} because the contribution of features to the predicted value is easily to observe. Moreover, it proves to be more consistent with human intuition~\cite{10.5555/3295222.3295230} than other interpretable machine learning techniques, such as \textit{LIME}~\cite{ribeiro2016should}. Also, the value of \textit{SHAP} can be directly used to add up to the actual predictive value, explaining the importance of the feature, which is consistent with our motivation for the subsequent visual design.

\subsection{Spatiotemporal Clustering}
\par Two challenges exist in tracking the evolution of RIS over long periods of time and large geographic scales. First, correctly delineating long time periods is critical to identifying potential patterns in the RIS evolution as reflected by business registration activity. Simply retrieving records by year may lead to duplicate analyses if no significant changes occur in the two subsequent periods. Second, even if we manage to divide the evolution of the RIS into different intervals, for each interval we are still confronted with a large amount of data that shows different spatial locations by the records of the enterprises belonging to that interval. It is important to abstract the characteristics of the corresponding enterprises while maintaining dynamic and multi-attribute details. In the following subsections, we perform spatio-temporal clustering in terms of \textit{time series segmentation} and \textit{geospatial clustering}.

\begin{figure*}[h]
    \centering
    \includegraphics[width=\textwidth]{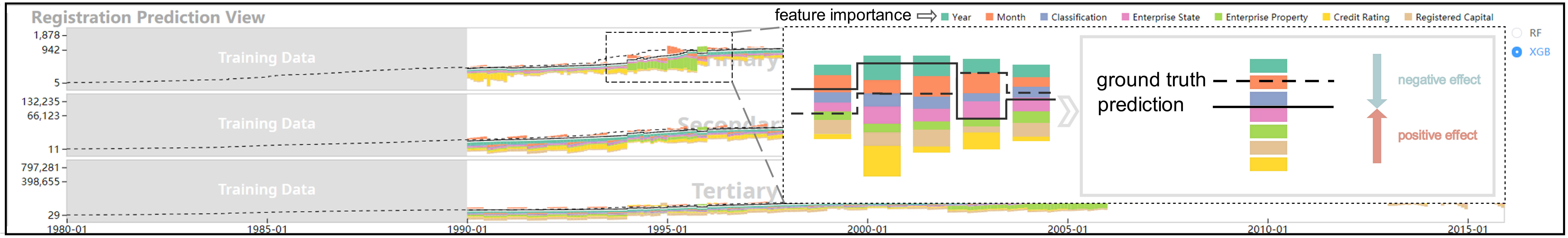}
          \vspace{-6mm}
    \caption{Three sub-charts from the top to the bottom represent primary, secondary and tertiary industry, respectively. Each sub-chart contains a dashed line representing the ground-truth value, a solid line representing the predicted value and stacked bars representing feature importances.}
       \vspace{-6mm}
    \label{fig:prediction}
\end{figure*}

\par \textbf{Partitioning of Time Series.} To address the first challenge, we formulate the partitioning of long periods of RIS evolution as a segmentation problem, which partitions long time series into segments that can be formulated as follows: given a time series $T$, generate the best representation such that the combined errors (which can be obtained by calculating the sum of the maximum error for all segments) is less than a user-specified threshold. Generating segments for such time series data is the key to an efficient and effective solution~\cite{keogh2004segmenting}. Notably, we take a top-down approach, considering every possible partition of the time series and segmenting it at the optimal location. The two subsegments are then tested to see if their approximation error is below a user-specified threshold. If not, the algorithm recursively splits the subsequence until the approximation error of all segments is below the threshold. Thus, the original time series registration is divided into several piecewise linear representations after time series segmentation.

\par \textbf{Geospatial Clustering.} To address the second challenge of the various enterprise activities within each segment, we analyze the business data from a cluster perspective rather than an individual perspective. The goal of clustering is to make the similarity between objects within the same category as large as possible and, conversely, the similarity between different categories as small as possible. It is worth noting that we tried to cluster individual enterprises belonging to different time periods using the division-based \textit{KMeans} algorithm and the density-based clustering algorithm \textit{DBSCAN}, respectively. However, they both exhibit specific advantages and disadvantages. For example, although \textit{KMeans} can achieve clustering results, the determination of $K$ is quite challenging because the location of enterprises is unpredictable. In addition, \textit{KMeans} is developed based on the centroid location of the aggregated clusters rather than the actual geographic locations of enterprises. On the other hand, \textit{DBSCAN} requires manual input of two parameters, namely $[Eps]$ and $[minPts]$, so the accuracy of the clustering results directly depends on the users' parameter selection. Also, the obtained clusters are spatially arbitrary shaped and are just sets of geographical points rather than aggregated clusters. In addition, \textit{DBSCAN} suffers from indistinguishable noisy data. For these reasons, we propose a hybrid algorithm based on \textit{DBSCAN} and \textit{KMeans} algorithms. First, the geographic locations of enterprises are clustered into several clusters using the density reachability of \textit{DBSCAN} according to the specific settings of the two parameters. The data in each cluster is taken as a new input. The centroid positions are then obtained using \textit{KMeans} to minimize the Sum of Squared Error (SSE) between the data points in each cluster and the centroids of the clusters they are in using the iterative aggregation of \textit{KMeans} and $K$ is set to $1$. In addition, we use \textit{KANN-DBSCAN}~\cite{li2019research}, to automatically find a stable interval of cluster number variation by generating candidate $[Eps]$ and $[MinPts]$ using the distribution characteristics of the dataset. We take the $[Eps]$ and $[MinPts]$ parameters corresponding to the minimum density threshold of this interval as the optimal parameters. Specifically, first, the corresponding $[Eps]$ candidate list can be obtained using \textit{KANN-DBSCAN}. Then, for the given $[Eps]$ candidate list, the number of neighboring objects can be calculated sequentially and the expected values of these $[Eps]$ are used as the $[MinPts]$ candidate list:
$
MinPts = \frac {1}{n}\sum _ {i=1}^ {n}P_ {i},
$
$P_i$ is the number of $[Eps]$ neighboring objects of the $i_{th}$ object, and $n$ is the total number of objects of the data. Finally, the list of these two parameters is given as input to \textit{DBSCAN}, and then the number of clusters generated with different parameter settings is obtained separately and quickly. The result can be considered stable when the number of clusters is the same three times in a row. After deciding the clusters, we need to further define the centroids of each cluster. To summarize, we first apply \textit{KANN-DBSCAN} to cluster the geographic locations of enterprises into several clusters, and then use \textit{KMeans} to find the centroid location of each cluster.

\section{Front-end Visualization}
\par \textit{RISeer} uses or enhances familiar visual metaphors so that domain experts can focus on analysis. We strictly follow the mantra of ``\textit{overview first, zoom and filter, then details-on-demand}''~\cite{shneiderman2003eyes}. As shown in \autoref{fig:pipeline}, we develop five main visualizations to help domain experts explore the evolutionary patterns of RIS at the overview, evolution pathway, and regional cluster levels. Specifically, we design the \textit{RIS Projection Overview} to obtain the overall distribution of RIS patterns over the entire periods (\textbf{R.1}). The \textit{Registration Prediction View} captures the correlation between the predicted registrations and ground truth over time, and understands the latent features driving the changes from the perspective of the time series forecasting models (\textbf{R.2 -- R.3}). The \textit{RIS Evolution View} explores the regional features and their evolutionary paths over time (\textbf{R.4}). The \textit{Spatiotemporal View} visually projects the identified regional clusters onto their corresponding locations to facilitate understanding map-specific RIS patterns such as mitigation, radiation, and aggregation (\textbf{R.4 -- R.5}). The \textit{Regional Comparison View} helps domain experts to make detailed comparisons in terms of geospatial indicators across multiple regions (\textbf{R.6}).

\subsection{RIS Projection View}
\par
\begin{figure}[h]
    \centering
        \vspace{-3mm}
    \includegraphics[width=0.5\textwidth]{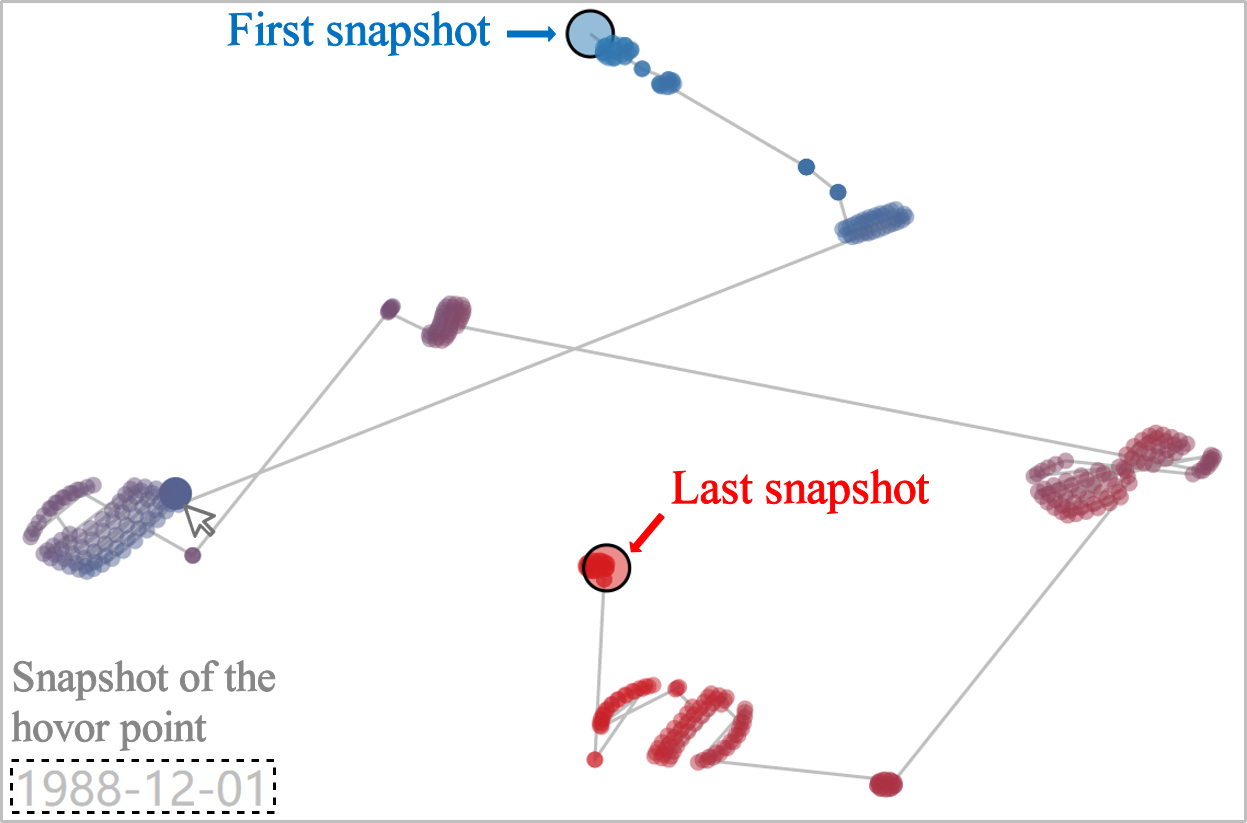}
    \vspace{-6mm}
    \caption{The RIS Projection View facilitates identifying the potential anomalous and ``clusters'' of RIS snapshots.}
    \label{fig:Projection}
    \vspace{-3mm}
\end{figure}

 The RIS Projection View helps to identify potential anomalies and ``clusters'' in RIS snapshots. Dimensionality reduction techniques such as \textit{t-SNE}, \textit{PCA}, and \textit{MDS} have been widely adopted to create low-dimensional representations that preserve local similarity to express neighborhood structure~\cite{10.1111:cgf.13417, 10.1109/TVCG.2016.2598838}. We follow the conventional practice of projecting all RIS snapshots for the entire period into two-dimensional space to see potential clusters and outliers. After discussions with experts, we use the following metrics, namely, \textit{enterprise classification code}, \textit{registered capital}, \textit{credit rating}, \textit{enterprise property}, and \textit{enterprise state}, to evaluate the RIS snapshots at the overview level. After obtaining the above metrics, we can obtain the corresponding feature vectors for each RIS snapshot. We use \textit{t-SNE} as the dimensional projection because ``\textit{it reveals meaningful insights about the data and shows superiority in generating two-dimensional projection}''~\cite{li2018embeddingvis}. Similar to ~\cite{van2015reducing}, as shown in \autoref{fig:Projection}, each RIS snapshot in two-dimensional space is represented by a point, and the color indicates the snapshot. The first and last snapshots are highlighted (red for the first snapshot and blue for the last) and a black curve connects all snapshots in chronological order. The user can hover over any particular RIS snapshot and a tooltip will display a detailed timestamp. The user can lasso any entity on the projection space for interaction and for further ``link + view'' analysis.

\begin{figure}[h]
    \centering
    \includegraphics[width=0.5\textwidth]{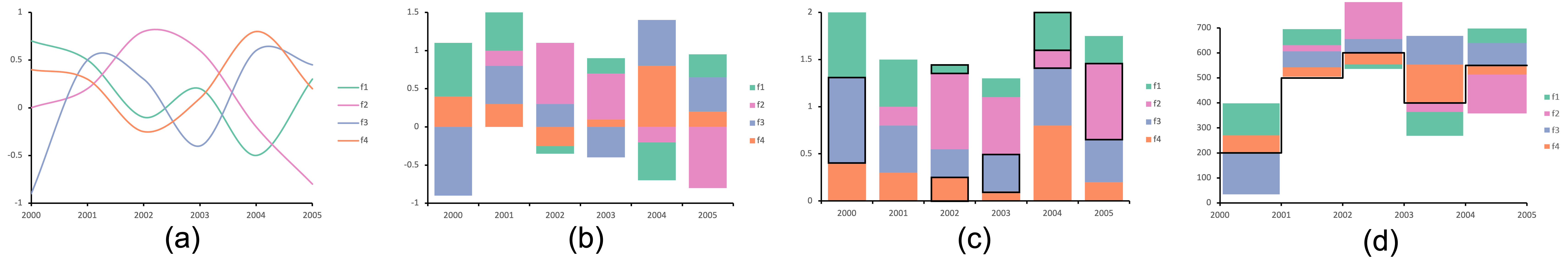}
    \vspace{-8mm}
    \caption{Design alternatives for the feature importance.}
    \label{fig:alternative}
    \vspace{-4mm}
\end{figure}

\begin{figure*}[h]
    \centering
    \includegraphics[width=\textwidth]{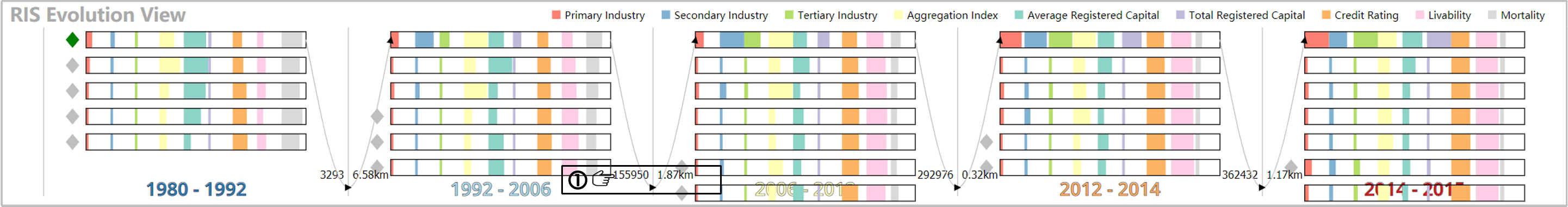}
          \vspace{-6mm}
    \caption{The RIS Evolution View helps domain experts conduct metric comparisons at the level of regions.}
    \label{fig:evolution}
        \vspace{-3mm}
\end{figure*}

\begin{figure*}[h]
    \centering
    \includegraphics[width=\linewidth]{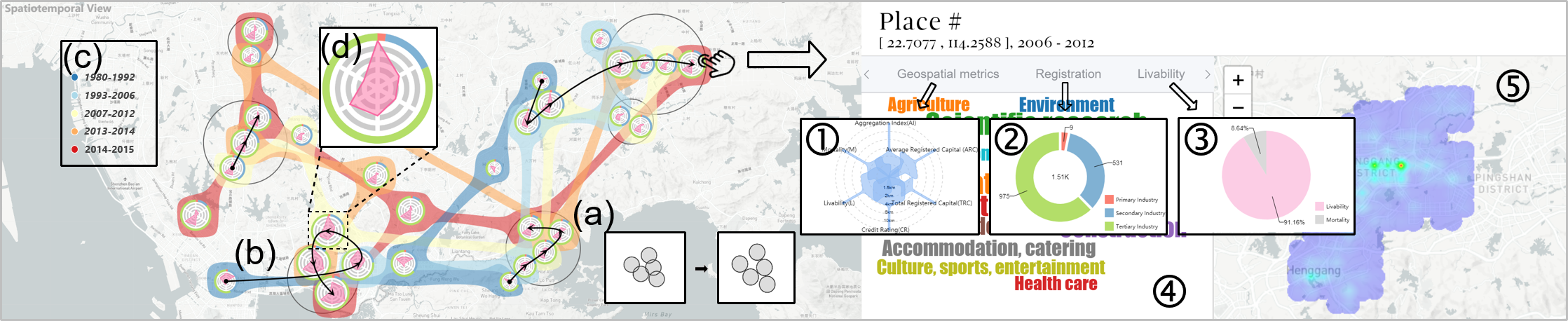}
          \vspace{-6mm}
    \caption{The Spatiotemporal View: (a) Packed circles based on physical simulator; (b) Regional evolution path; (c) Five-colored bubble sets indicate clusters at different time periods; (d) The glyph represents an overview of regional information. After clicking one glyph, more information will be shown in a separate panel: (1) Geospatial distribution of six metrics in glyph; (2) Registration information; (3) Enterprise livability; (4) Business category, and (5) Heat map represents actual geographical coverage of the regional cluster.}
    \label{fig:Spatiotemporal}
          \vspace{-6mm}
\end{figure*}

\subsection{Registration Prediction View}
\par The Registration Prediction View (\autoref{fig:prediction}) helps experts understand the detailed associations between RIS influencing factors and registration amounts based on several representative time-series forecasting models. It consists of three subplots, arranged from top to bottom, representing business activities in the primary, secondary and tertiary sectors, respectively. In each subplot, the dotted line represents the ground truth registration of enterprises. The solid black line represents the predicted business activity compared to the ground truth volume. Thus, users can observe the gap that represents the difference between the actual and the predicted value at different timestamps. In addition, the importance of the $7$ features is calculated, normalized, and represented with different classification colors stacked on top of each other. In each snapshot, the glyphs of feature importance are placed along the Y-axis. Each bar in the stacked glyphs represents a feature. Bars stacked above the predicted values represent features with negative effects, pushing the predicted value down~\cite{10.1145/3411764.3445083}. The features stacked below the predicted value are positive, pushing the predicted value higher. The user can select other models on the right side for comparison.

\par \textbf{Design Alternatives.} We consider several alternative designs as shown in \autoref{fig:alternative}. \autoref{fig:alternative}(a) simply uses a multilinear graph to represent the importance of features over time. When the number of features is large, visual clutter can be severe. To alleviate this problem, we propose \autoref{fig:alternative}(b), which replaces the polylines with stacked bars. However, it cannot show the continuous change of features because the order of different features is not fixed at each timestamp. \autoref{fig:alternative}(c) is an improvement on the second one. The height of the bars encodes the absolute feature importance value, and the bars corresponding to negative features are highlighted with strikes. In this design, the order of features is fixed, allowing to track the changes in feature importance over time. However, it mixes positive and negative features and does not relate the predicted values to the importance of the features. We finally chose \autoref{fig:alternative}(d) because it better distinguishes between positive and negative features and embeds the predicted values in the design.

\subsection{RIS Evolution View}
\par Although the above RIS Projection View conveys additional information about clusters and potential outliers in all RIS snapshots, it is still unclear how the RIS patterns evolve at the regional level. In addition, domain experts often need to explore the properties of each regional cluster in order to compare and evaluate them in one or more temporal snapshots, which can help them to assess the performance of RIS. Therefore, based on discussions with experts, we adopt and standardize the following indicators for a uniform comparison: 1) \textit{number of primary industries}; 2) \textit{number of secondary industries}; 3) \textit{number of tertiary industries}; 4) \textit{aggregation index} ($AI(x)$), which is calculated by the Coefficient of Variation~\cite{everitt1998}, to measure the similarity of the distribution of indicators as follows:
$
CV(x) = \frac{\sqrt{\frac{1}{N}\sum_{i=0}^N{(x_i - \mu)}^2}}{\mu},
$
and $AI(x) = \frac{1}{CV(x)}$; 5) \textit{average registered capital}; 6) \textit{total registered capital}; 7) \textit{credit rating}; 8) \textit{livability}, i.e., the number of surviving enterprises divided by the total number of enterprises; and 9) \textit{mortality}, complementary to livability.

\par Inspired by \textit{Lineup}~\cite{gratzl2013lineup}, we design a RIS Evolution View to help domain experts to compare metrics at the regional level. Notably, as shown in \autoref{fig:evolution}, we present the metric values for each region cluster as a combined bar, where the length of a single colored bar indicates the normalized metric value for the corresponding region cluster. We rank the region clusters generated in all consecutive snapshots. The ranking is based on the value of a specific metric. When a region cluster is clicked, all ``identical'' region clusters across time snapshots are highlighted. We determine whether a regional cluster is identical between two subsequent snapshots based on the amount of overlaps, and we only depict the evolutionary path between two subsequent ``identical'' region clusters with the largest number of overlaps. In the two subsequent snapshots, we place an axis with two different scales along the left and right sides of the axis. The left axis represents the scale of the number of enterprises that transition from one regional cluster to another in the next period. The right axis represents the scale of the distance between the two centroids of the regional clusters in the two subsequent snapshots. For example, as shown in in \autoref{fig:evolution}(1), these two figures indicate that $155,950$ enterprises in one regional cluster between $1992$ and $2006$ stay in the other regional cluster between $2006$ and $2012$, while the distance between the centroids of the two regional clusters is $1.87km$ each. Intuitively, a small value indicates that the main distribution of enterprises in the early snapshot is largely consistent with the distribution of enterprises, including the existing enterprises in the early snapshot and newly established ones in the later snapshot.

\begin{figure*}[h]
    \centering
    \includegraphics[width=\linewidth]{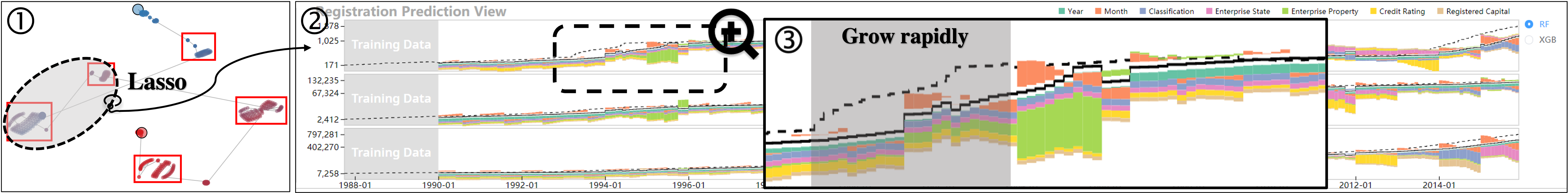}
          \vspace{-6mm}
    \caption{The RIS projection view projects Shenzhen's enterprises' registration data into several clusters.}
    \label{fig:pp}
          \vspace{-6mm}
\end{figure*}

\subsection{Spatiotemporal View}
\par We apply the classical map-centric visual exploration approach to represent clusters of enterprise geographic locations. As shown in \autoref{fig:Spatiotemporal}, we design radar-like glyphs at the top of the map and place them in the corresponding geographic regions of the regional clusters generated by the geospatial clustering described above. The indicators represented by the axes in the radar-like glyphs are the same as those in the RIS evolution view, namely, \textit{aggregation index}, \textit{average registered capital}, \textit{total registered capital}, \textit{credit rating}, \textit{livability}, and \textit{mortality}. We did not choose a bar chart because we need to have the radius of the glyphs represent the number of enterprises in the cluster. In addition, in the Regional Comparative View (see Section 6.5), we need to compare the same indicators between two or three clusters, and radar-like glyphs are easier to perceive than bar charts for comparison.

\par Note that we place all the regional clusters generated by \textit{DBSCAN} and \textit{KMeans} on the map at different time intervals, inevitably causing visual clutter and overlap problems. Therefore, we use a force-directed layout to pack those overlapping clusters~\cite{collins2003circle}. As shown in \autoref{fig:Spatiotemporal}(a), a physics-based simulator will be used to find the optimal circle positions by 1) optimizing the distance between the circle and the force center, 2) attracting each other slightly, and 3) avoiding overlap. We warp the packed circles with an additional black ring (\autoref{fig:Spatiotemporal}(a)), indicating that these groups of regions have some kind of unavoidable overlap. To emphasize the representation of region groups from a particular time snapshot and inspired by~\cite{alsallakh2016state}, we borrow the \textit{Bubble Sets} technique~\cite{collins2009bubble} to provide continuous boundary contours that allow us to examine scenarios with semantically important spatial organization and important set membership relationships for enterprises belonging to the snapshot. Notably, as shown in \autoref{fig:Spatiotemporal}(c), the five colored bubble sets represent the geospatial locations in the five snapshots. We also provide regional evolution paths in the form of black curves that connect clusters of regions in different snapshots in chronological order (\autoref{fig:Spatiotemporal}(b)).

\par To elaborate more detailed information about region-specific clusters, we provide a separate panel (\autoref{fig:Spatiotemporal}(1 -- 5)) showing the \textit{geospatial metrics} (\autoref{fig:Spatiotemporal}(1)), \textit{registration} (\autoref{fig:Spatiotemporal}(2)), \textit{livability} (\autoref{fig:Spatiotemporal}(3)), and \textit{business category} (\autoref{fig:Spatiotemporal}(4)). The actual geographical coverage of the regional clusters is displayed as a heat map (\autoref{fig:Spatiotemporal}(5)), where domain experts can observe the most dense points and the scale of distribution.

\subsection{Regional Comparative View}
\par While it is crucial to get a general impression and assess the overall distribution of indicators for regional clusters on a map, experts also want to understand how they differ from each other and whether one specific regional cluster is better than another. That is, multiple regional clusters should be placed in the context of each other in order to thoroughly compare relevant information and differences across regions simultaneously. When users find several ideal candidate regional clusters through the Spatiotemporal View, they can add them in for detailed comparison. Inspired by \textit{Skylens}~\cite{zhao2017skylens}, we design a Regional Comparative View that allows users to take a closer look at the differences between several regional clusters (\autoref{fig:pipeline}(E)). Specifically, we invite a radar plot-based glyph to perceive the factor values of different regional clusters. For our specific scenario, we enhance the traditional radar plot in several ways. First, the axes show the same indicators as in the spatiotemporal view. However, since most of these indicators are numerical values rather than distributions, it is not feasible to represent the detailed distribution of each indicator at the regional cluster level. Inspired by the RIS agglomeration phenomenon, we decide to calculate the indicator distribution based on geographic agglomeration. In other words, each glyph is represented as a series of rings, representing geographic distances. Specifically, after analyzing the location distribution, we arrange the distances from the cluster centroid into segments, i.e., $0$ -- $1.5$km, $1.5$ -- $2$km, $2$ -- $4$km, $4$ -- $6$km, and $6$ -- $10$km. Thus, each indicator can be restructured according to distance segments in an attempt to answer questions such as ``\textit{is there an industrial agglomeration in such regional clusters?}'' To mitigate the distortion caused by the visual overlap of different metrics, the user can highlight a particular metric by hovering over it. The differences between two regional clusters are described by radar glyphs illustrating the indicator distributions in the two corresponding regional clusters for comparison. In addition to comparing selected regional clusters, the growth rates for each industry along the evolving RIS paths provide insight into regional development, especially if the region is expanding its geographic size. As shown in \autoref{fig:pipeline}(E), we plot the growth rate of business registration across all snapshots using box plots, with each representing one region of a particular RIS path. Multiple box plots can be aligned to facilitate comparative analysis of growth rates across regions in a single snapshot.

\subsection{Interaction Among the Views}
\par Rich interactions are integrated to facilitate exploring and assessing RIS patterns. Users can select regional clusters of interest in either RIS Evolution View or Spatiotemporal View, and \textit{RISeer} will highlight the corresponding information in other views. The time range in Registration Prediction View can be adjusted interactively to highlight specific periods of RIS evolution. The coordinated interaction between comparable regional clusters facilitates the inspection of a variety of information at different granularities and other contexts. Users can also hover the mouse over the path of a bubble set and the covered regional clusters will be highlighted for visualization.

\section{Evaluation}
\subsection{Case \uppercase\expandafter{\romannumeral1}: Shenzhen Leading China's Economic Miracle}
\par Below are a series of activities that occurred in E1 -- 2 while exploring Shenzhen business registration records over the past $30$ years using \textit{RISeer}. In particular, Case 1 uncovers RIS evolution patterns and key observations of Shenzhen that lead China's economic miracle.

\begin{figure}[h]
    \vspace{-3mm}
    \centering
    \includegraphics[width=\linewidth]{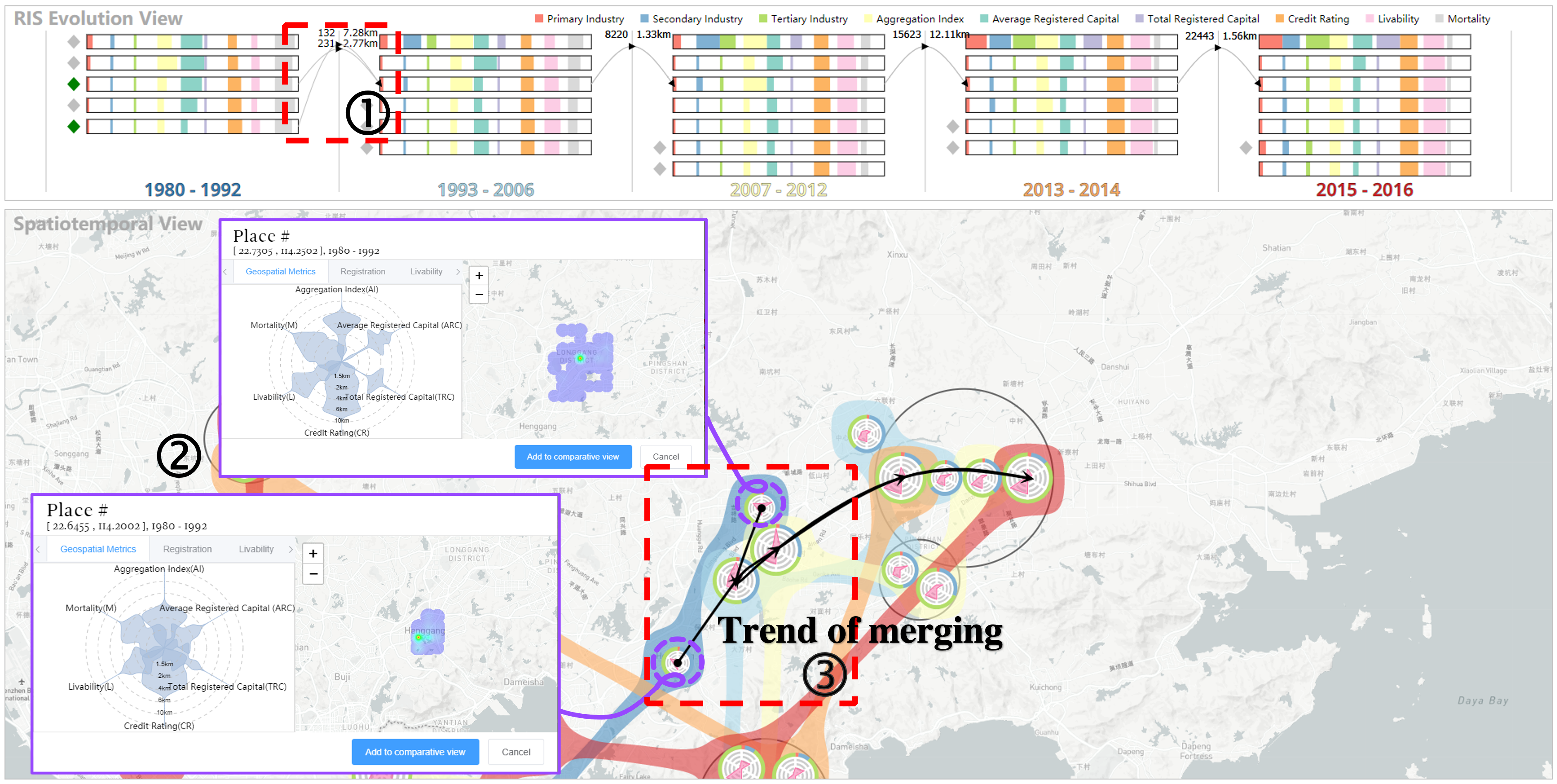}
          \vspace{-6mm}
    \caption{Identifying two country-level districts were merged into one district by observing two RIS evolving and merging paths.}
    \label{fig:st}
        \vspace{-3mm}
\end{figure}

\begin{figure*}[h]
    \centering
    \includegraphics[width=\linewidth]{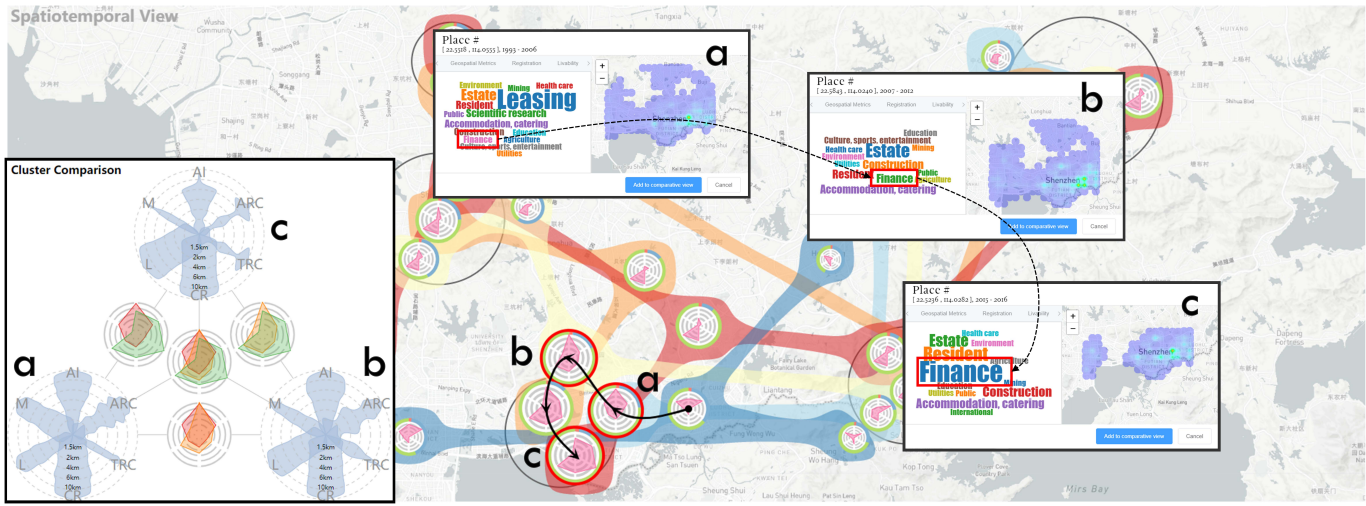}
          \vspace{-6mm}
    \caption{The financial sector in the selected regional cluster (i.e., Futian District, Shenzhen) is gradually occupying a dominant position over time.}
    \label{fig:comparison}
          \vspace{-6mm}
\end{figure*}

\par After experts loaded a total of $1.1$ million in business registration activity in Shenzhen over the past $30$ years into \textit{RISeer}, the RIS Projection View breaks down Shenzhen business registration data by month. It describes the corresponding business registration data for each month based on a feature vector consisting of \textit{enterprise classification code}, \textit{enterprise state}, \textit{enterprise property}, \textit{credit rating} and \textit{registered capital}. In other words, a feature vector is generated for the business registration data in Shenzhen in the monthly snapshot. After projecting the high-dimensional feature vectors onto a two-dimensional plane, experts observed several different clusters (\autoref{fig:pp}(1)). E2 then lassoed this view and determined that each cluster corresponded to a specific continuous period. The experts were quite interested in those projected observations with large distance spans, so E2 directly performed a larger lasso operation on this view (\autoref{fig:pp}(2)). As shown in \autoref{fig:pp}, after the lasso operation, the registration prediction view predicts the registration data of enterprises for the corresponding time horizon, including ground truth, model predictions, and feature importance. As shown in \autoref{fig:pp}(3), E1 observed that the number of enterprises in Shenzhen grows very rapidly from $1993$ to $1998$. E1 also found that at the beginning of this rapid rise, the difference between the model prediction and the ground truth is large, but this difference slowly decreases over time. Therefore, E1 believes that some different events occurred during this period and that this pattern of change is new to the model and will take some time to learn. E2 further noticed that the feature importance captured by the model during this period shows a sharp increase in the contribution of features of enterprise attributes in the primary sector, while the changes in the secondary and tertiary sectors are the result of a combination of features. E2 then moved to the RIS Evolution View, focusing on the paths corresponding to that period. As shown in \autoref{fig:st}, after clicking on multiple evolutionary paths, the expert found a tendency for two of them to merge during this period (\autoref{fig:st}(1)). The expert then clicked on each cluster glyph to view the geographic coverage through the heat map and found that the starting areas of the two paths were in \textit{Henggang District} and \textit{Longgang District} (\autoref{fig:st}(2)). The expert added that before $1993$, \textit{Henggang} and \textit{Longgang} were two township-level districts in Shenzhen, respectively, and after $1993$, they were merged into \textit{Longgang District} due to administrative planning (\autoref{fig:st}(3)). Moreover, E1 mentioned that the rapid growth of newly registered enterprises in this area has led to the relocation of cluster centroids, ``\textit{the rural shareholding cooperative system has been implemented in this region.}'' That is, the nature of the primary sector has changed. This also explains the predictive model believes that the feature of enterprise property was the main contributing factor in this period. Experts also mentioned that the ``\textit{`Three-Plus-One' trading-mix strategy (i.e., customized production and compensatory trade in materials, designs, or samples) stimulated the development of the local economy, leading to a rapid growth in the number of regional businesses}.``

\par In addition to observing the above phenomena of RIS path evolution and merging, experts would like to explore how the RIS evolves along the RIS paths. Notably, as shown in \autoref{fig:comparison}(a -- c), E1 relocated a RIS path, selected several regional clusters in chronological order along the path, and added them to the Cluster Comparison View for further comparison. E1 witnessed that over time, the \textit{registered capital}, the \textit{livability}, and the \textit{credit rating} of regional enterprises are increasing, which indicates a positive momentum of industrial development in the region. By looking closely at the information of each cluster, experts found that the financial industry in the region (i.e., Futian District, Shenzhen) is becoming dominant over time. This development trend also reflects the spirit of the ``\textit{Shenzhen Special Economic Zone Development Plan}'', promulgated in $1982$, which clearly defines Futian as a commercial, financial and administrative center.

\begin{figure*}[h]
    \centering
    \includegraphics[width=\linewidth]{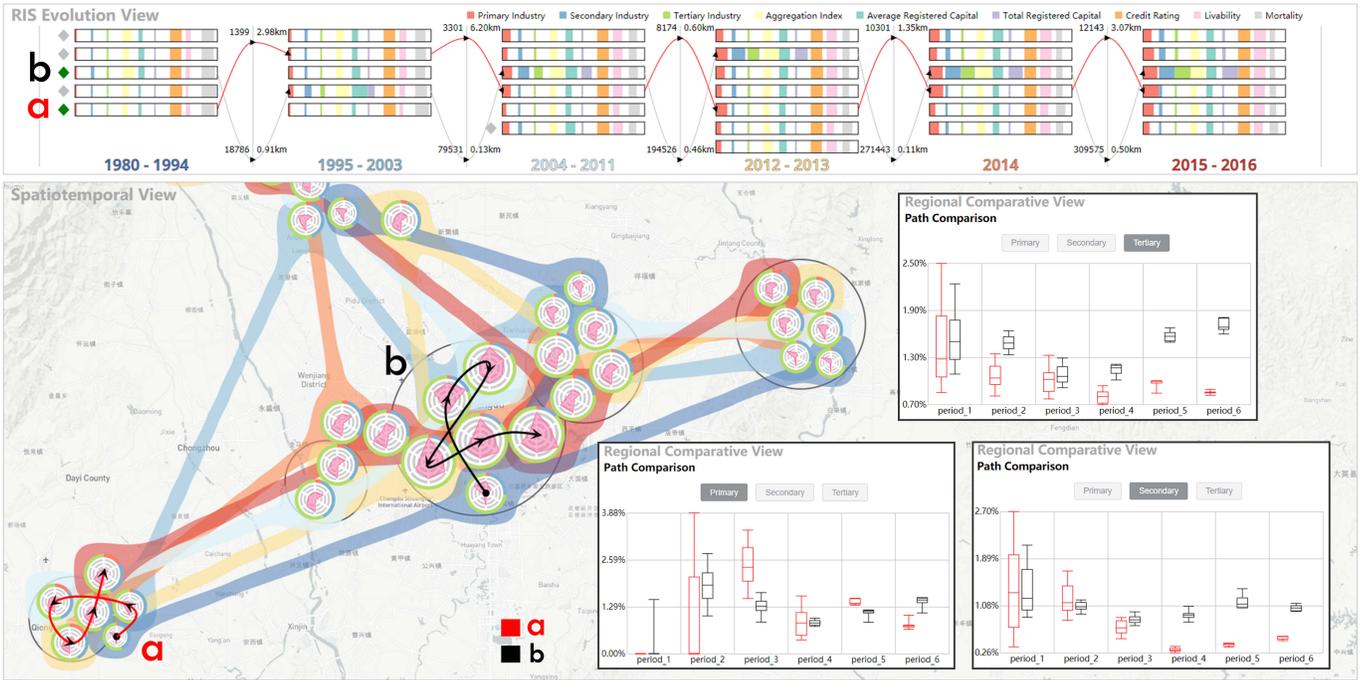}
          \vspace{-8mm}
    \caption{Two paths: (a) is based on the primary industry development and (b) is based on the secondary/tertiary industries.}
    \label{fig:case2-1}
          \vspace{-6mm}
\end{figure*}

\subsection{Case \uppercase\expandafter{\romannumeral2}: Development Comparison Between Wuhou District and Qionglai District in Chengdu}

\par
\begin{figure}[h]
    \centering
    \includegraphics[width=\linewidth]{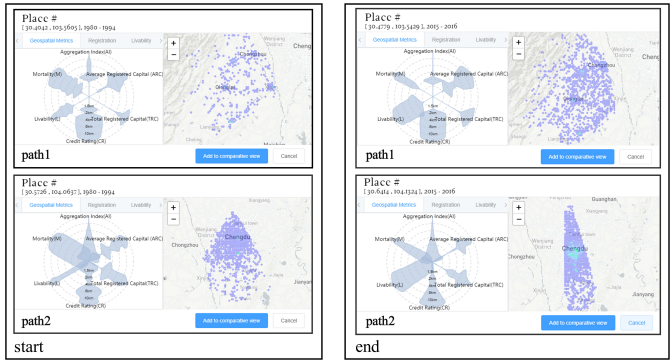}
          \vspace{-6mm}
    \caption{Comparing the starting/ending regional clusters for two paths.}
    \label{fig:case2-2}
          \vspace{-6mm}
\end{figure}

 This case documents our experts' use of \textit{RISeer} to study business registration activities in Chengdu. E1 and E2 first turned their attention to the RIS Evolution View and interactively examined the evolutionary patterns of the different paths. They found that one of the paths (path1 (\autoref{fig:case2-1}(a))) is mainly based on the development of the primary sector. In contrast, the other path (path2 (\autoref{fig:case2-1}(b))) is mainly based on the development of the secondary and tertiary sectors. Based on the geographical location information related to the regional cluster provided by the Spatiotemporal View, the experts witnessed that the enterprises covered by path1 are mainly located in Qionglai (a country-level city affiliated with Chengdu). As shown in \autoref{fig:case2-1}, in the RIS Evolution View, the experts further observed that the primary industry of path1 has been proliferating since $2004$. Based on the above information, experts investigated and studied the relevant periods and locations and found that in order to balance urban and rural development, Chengdu established a model of arable land protection fund system in $2008$, and Qionglai was the key area of arable land protection. Especially, ``\textit{after the implementation of the farmland protection fund system, the arable land area in Qionglai has increased year by year}'', and corresponding agricultural output value has also been increasing. These policies have greatly stimulated the development of the primary industry. Meanwhile, through the heat map distribution of the clusters corresponding to path2 (\autoref{fig:case2-2}), the experts could see that the enterprises participating in path2 were mainly in the urban area of Chengdu. They also found that the number of enterprises in path2 increased significantly after entering period2. In particular, the number of enterprises that existed in $2004$ -- $2012$ was twice as many as in $1995$ -- $2003$. The secondary industry grew rapidly, ``\textit{the deepening industrial reforms in Chengdu are due to the impact of large-scale development policy in the western region.}''

\par The experts also found different growth rates for the two paths from the Path Comparison View. It can be seen that the growth rate of the primary sector for path1 is generally higher than that of path2. On the contrary, the growth rates of the secondary and tertiary sectors of path2 are significantly higher than those of path1 after $2004$ (\autoref{fig:case2-1}). In fact, the regions corresponding to path2 have entered a rapid development phase after $2004$. By comparing the starting aggregation information of the two paths (\autoref{fig:case2-2}), the experts found that the starting aggregation of path2 is more uniform than that of path1 in terms of the distribution of various indicators. The aggregation index of path2 is also relatively higher than that of path1. This conclusion can also be verified by the heat map corresponding to cluster, where the experts determined that the geographical distribution of the starting cluster of path1 is more discrete. In contrast, the geographic distribution of the starting cluster of path2 is more dense. Likewise, experts can find that the average registered capital and total registered capital of the end cluster of path2 are significantly higher than those of the end clusters of path1. This is due to the fact that path2 develops mainly secondary and tertiary industries, while path1 develops primary industries.

\section{Discussion and Limitation}
\par We conducted one-hour semi-structured interviews with E1 -- E4 to verify the efficacy of our methodology to study RIS dynamics.

\par \textbf{System Performance.} All experts appreciated the ability of our system to support exploration of RIS evolution patterns in large volumes of business registration data. ``\textit{RISeer is very cool because it provides a highly interactive exploration to detect regional clusters and time periods,}'' said E1. \textit{RISeer} can examine a large number of enterprises activities at different levels of granularity, which can ``\textit{substantially reduce the amount of work consumed by traditional data exploration.}'' E2 liked the design of Spatiotemporal View because he had tried to present the location of all enterprises on a map, but there was a serious visual clutter. \textit{RISeer} embeds spatiotemporal factors into the map, which ``\textit{resolves my concerns when doing customized exploration and decision making.}'' E3 commented that ``\textit{it inspired me when it provided some visual cues to remind me of some old messages and strategies.}'' E4 said that \textit{RISeer} has the potential to be deployed to real-world scenarios and help RIS researchers with retrospective analysis. E1 and E2 liked the idea of using predictions to examine the underlying data. E3 echoed his opinions, saying ``\textit{the projection overview is very intuitive because it fits my intuition about the evolving RIS in Shenzhen.}'' We conducted a user-centric design process that involved experts in the initial design phase so they soon became familiar with the visual encoding.

\par \textbf{Contributions Over Previous Work.} E3 indicated that \textit{RISeer} can effectively ``\textit{extract the main features and factors that drive regional economic and industrial restructuring}''. Experts can identify the patterns of regional development from the static historical data. If the industrial structure has changed significantly in a certain period, \textit{RISeer} can give direct hints to capture this signal in time. In addition, traditional RIS analysis aggregates the companies involved on a map based on their geographic locations and divides them in fixed-size blocks. Nevertheless, such fixed divisions can undermine the integrity of company aggregation. \textit{RISeer} can effectively display the phenomenon of industrial aggregation, radiation and migration, allowing users to select specific areas and track RIS changes within that area.

\par \textbf{Generalizability and Scalability.} We discussed with the experts which parts of \textit{RISeer} could be used directly for other scenarios and which parts needed customization. They said that the system is already very versatile and can be applied to any region if the data is pre-processed beforehand. Also, other economic data, such as GDP, can be easily integrated into \textit{RISeer}. This work has the potential to be applied to a variety of usage scenarios that leverage ``alternative data'' to e.g., understand economic status, guide investment strategy and find a market edge. Regarding the scalability, \textit{RISeer} handles millions of data records for a single city by abstracting to regional clusters. Therefore, scalability issues for front-end visualization are minimal. The bottleneck of algorithmic complexity lies in clustering. When the data is huge and complex, the density of locations extracted each time may not be consistent, which may not produce satisfactory results.

\par \textbf{Limitation}. First, there is a lag in updating enterprise information due to occasional lax enforcement and incomplete filling. Our data may inevitably contain such outdated information, which may bias the objective evaluation and analysis of the current state of the RIS. Second, we divided all enterprises into three significant industries without further subdivision. This is due to the wide variety of subsectors within the three significant industries and the different criteria in each country. However, this also prevents our system from performing a more careful analysis of the substructures within the sectors. Third, although we use multivariate time-series forecasting models to predict the activities of enterprises, registration information may have different criteria at an early stage, resulting in some missing data attributes and inaccurate results. And we classified the target period based on the changes in the overall number of business registration activities, which may ignore other features. We could try the same time series segmentation algorithm in a higher dimensional space, or support users to segment the data themselves. Fourth, among the nine metrics abstracted from regions, the livability and extinction rates of enterprises are time-varying and should not be put together with the other indicators. Moreover, different color encoding in different views may cause visual confusion.

\section{Conclusion and Future Work}
\par We propose a visual analytics approach to exploring and analyzing the evolutionary dynamics of RIS revealed by large-scale regional business registration activities. We first summarize the design requirements through in-depth discussions with domain experts. The proposed \textit{RISeer} facilitates the examination and exploration of key time periods, dynamic correlation between predictions and ground truth by feature importance, and regional clusters and their spatiotemporal 
properties. Two case studies and expert interviews demonstrate the effectiveness of our approach. Currently, \textit{RISeer} is more like a confirmatory analysis tool for observing and analyzing the changes in the existing RIS. We plan to extend our approach to a broader scope, capturing dynamic RIS evolution patterns using time series forecasting models better suited to change and city-level comparisons.

\acknowledgments{
We are grateful for the valuable feedback and comments provided by the anonymous reviewers. This work is partially supported by the Key Projects of Shanghai Soft Science Research Program from the Science and Technology Commission of Shanghai Municipality (No. 22692112900, 21692108700).}

\bibliographystyle{abbrv-doi}

\balance
\bibliography{template}
\end{document}